\documentclass[a4paper,11pt]{article}
\usepackage{pos}
\usepackage{mathtools}
\usepackage{etoolbox}
\usepackage{tikz}
\usetikzlibrary{shapes.geometric}
\usepackage{slashed}
\usepackage{subcaption}
\usepackage{braket}
\usepackage{listings}
\usepackage[dvipdfmx]{colortbl}
\usepackage{xcolor}
\usepackage{appendix}
\usepackage{hyperref}
\usepackage{enumitem}
\usepackage{bm}
\usepackage{diagbox} 
\usepackage{listings}
\usepackage{float}
\usepackage{multirow}
\usepackage{lipsum}
\usepackage{array}
\usepackage{float}
\usepackage{bm}

  \DeclareMathOperator{\Tr}{Tr} 
 \newcommand{\be}{\begin{equation}}
\newcommand{\ee}{\end{equation}}
\newcommand{\bea}{\begin{eqnarray}}
\newcommand{\eea}{\end{eqnarray}}

\newcommand{\bi}{\begin{itemize}}
\newcommand{\ei}{\end{itemize}}

\begin{document}

\title{
Study of multi-particle states with tensor renormalization group method
}

\author*[a]{Fathiyya Izzatun Az-zahra}
\author[a]{Shinji Takeda}
\author[b,c]{Takeshi Yamazaki}
\affiliation[a]{Institute for Theoretical Physics, Kanazawa University,\\ Kanazawa 920-1192, Japan}
\affiliation[b]{Institute of Pure and Applied Sciences, University of Tsukuba,\\ Tsukuba 305-8571, Japan}
\affiliation[c]{Center for Computational Sciences, University of Tsukuba,\\ Tsukuba 305-8577, Japan}
\emailAdd{fathiyya@hep.s.kanazawa-u.ac.jp}
\emailAdd{takeda@hep.s.kanazawa-u.ac.jp}
\emailAdd{yamazaki@het.ph.tsukuba.ac.jp}
\abstract{We investigate the multi-particle states of the (1+1)-dimensional Ising model using a spectroscopy scheme based on the tensor renormalization group method. We start by computing the finite-volume energy spectrum of the model from the transfer matrix, which is numerically estimated using the coarse-grained tensor network. We then identify the quantum number and momentum of the eigenstates by using the symmetries of the system and the matrix elements of an appropriate interpolating operator. Next, we plot the energy for a particular quantum number and momentum as a function of system size to identify the number of particles in the corresponding energy eigenstates. With this method, we obtain one-, two-, and three-particle states. We also compute the two-particle scattering phase shift using L\"uscher's formula as well as the wave function approach, and compare the results with the exact prediction.
}
\FullConference{The 42nd International Symposium on Lattice Field Theory\\
TIFR, Mumbai, India\\
2- 8 November, 2025 }
\maketitle

\section{Introduction}
The investigation of multi-particle interactions using lattice field theory has been progressively done over the last decades.
In this framework, one computes the finite-volume energy and connects the result with the infinite-volume observables using a quantization condition \cite{Luscher:1986pf}.   
While the Monte Carlo algorithm is a powerful method for this calculation, it has some technical problems. For example, a lattice with a sufficiently large temporal extent is needed to suppress the contamination from higher excited states, and large statistical noise appears when extracting higher excited states.
Another method based on the tensor network
has been introduced in \cite{Itou:2023img,Itou:2024psm} for the Hamiltonian approach, 
and in \cite{PhysRevD.110.034514,Az-zahra:2024pqa} for the Lagrangian approach.
We focus on the latter, whose advantages are that a lattice with a small time extent is sufficient to extract the spectrum, and there is no statistical error because the scheme is deterministic.
However, the error of this scheme comes from the coarse-graining
using a tensor renormalization group algorithm \cite{PhysRevB.86.045139}.
In \cite{PhysRevD.110.034514,Az-zahra:2024pqa}, a square tensor network is used
for the scheme, yielding the low-lying energy results with small errors.
However, the errors drastically increase for higher excited states.
In the present work, we change the coarse-graining strategy
so that the higher excited states, which potentially correspond to the multi-particle states, can be extracted.
We coarse-grain the tensor network with several sizes in time direction and extract energy from the estimation of transfer matrix constructed from the coarse-grained tensors.
The quantum number and momentum of energy eigenstates are identified
using matrix elements represented as impurity tensor network.
By examining the energy at a fixed quantum number and momentum as a function of system size,
we identify the one-, two-, and three-particle states.
The scheme to compute the two-particle state wave function is also introduced.
Lastly, we compute the two-particle scattering phase shift 
from the finite-volume energy as well as from the wave function, 
and check the consistency of both approaches with the exact prediction.
For the demonstration, we apply the scheme to the (1+1)d Ising model.
\section{Spectroscopy scheme with transfer matrix}
We begin by reviewing the spectroscopy scheme introduced in \cite{PhysRevD.110.034514,Az-zahra:2024pqa} for the (1+1)d Ising model.
The partition function of the model is given by 
\be\label{eq:partition_function}
Z=\sum_{\{s=\pm 1\}} e^{\frac{1}{T}\sum_{\bm{r}\in \Gamma}\sum_{\mu=0}^{1}s(\bm{r}+\hat{\mu})s(\bm{r})} 
\ee
where $T$ is the temperature, $s(\bm{r})$ is the spin at $\bm{r}\in\Gamma$, $\hat{\mu}$ is the unit vector for the time ($\mu =0$) and space ($\mu=1$) directions,
and $\Gamma$ is a two-dimensional square lattice defined by
$\Gamma=\{\bm{r}=(t,x)|t=0,1,2,\ldots,L_{\rm t}-1\text{~and~} x=0,1,2,\ldots,L_{\rm s}-1\}$.
Here, $L_{\rm t}$ and $L_{\rm s}$ are the system size in time and space direction, respectively.
For system under periodic boundary condition (PBC), the partition function can be written as  
\be\label{eq:partition_function_TM}
Z=\Tr[{\cal T}^{L_{\rm t}}]
\ee
where $\cal{T}$ is the transfer matrix, which is explicitly given by
\be
\mathcal{T}_{S'S}=\left(\prod_{x=0}^{L_{\rm s}-1}\exp[{\frac{1}{T} s(t+1,x)s(t,x)}]\right)
\times
\left(\prod_{x=0}^{L_{\rm s}-1}\exp\left[{\frac{1}{2T}s(t+1,x+1)s(t+1,x)}\right]
\exp\left[{\frac{1}{2T}s(t,x+1)s(t,x)}\right]\right).
\ee
The term inside the first and second parentheses represent the interaction of spin fields in the time 
and space direction, respectively,
while 
$S'=\{s(t+1,x)|x=0,1,2,\ldots,L_{\rm s}-1\}$ and 
$S=\{s(t,x)|x=0,1,2,\ldots,L_{\rm s}-1\}$ 
denote the spins configurations at time $t+1$ and $t$.
\begin{figure}[t!]
\centering
\begin{subfigure}[b]{0.25\textwidth}
\centering
\includegraphics[width=3cm,height=3.5cm]{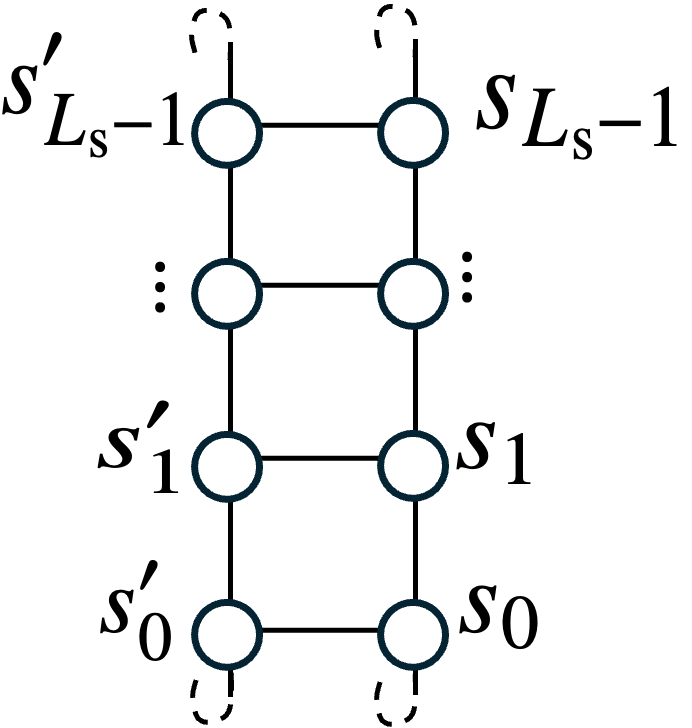}
\caption{}
\label{sfig:transfer_matrix}
\end{subfigure}
\begin{subfigure}[b]{0.25\textwidth}
\centering
\includegraphics[width=3cm,height=3.5cm]{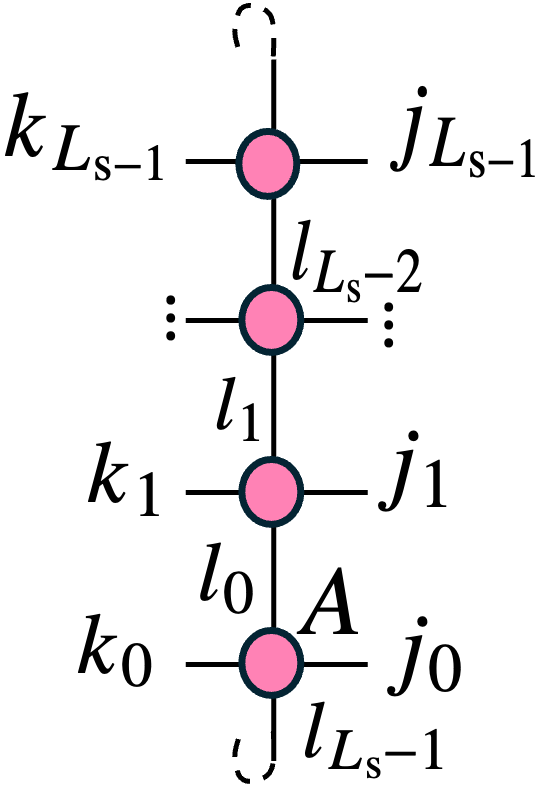}
\caption{}
\label{sfig:tensor_network}
\end{subfigure}
\caption{
A graphical image of (a) transfer matrix ${\cal T}_{S'S}$,
(b) one time slice tensor network ${\cal A}_{kj}$.
}
\label{fig:pict_TN}
\end{figure}
In Fig.~\ref{sfig:transfer_matrix}, we show the diagram of transfer matrix, where $s'_x, s_x$ notations are $s(t+1,x)$ and $s(t,x)$, respectively.
The energy spectrum can be obtained from the eigenvalues decomposition (EVD) of transfer matrix
\be\label{eq:evd_TM}
{\cal T}_{S'S}=\sum_{a=0}^{2^{L_{\rm s}}-1}U_{S'a}\lambda_a U_{aS}^{\dagger}
\ee
where $\lambda_a=e^{-E_a}$ is the eigenvalue corresponding to energy $E_a$,
and $U$ is the unitary matrix of the eigenvectors.
Accordingly, the energy gap of the system, $\omega_a=E_a-E_{\Omega}$, can be obtained from 
$\omega_a=\log\left(\frac{\lambda_0}{\lambda_a}\right)$ for $a=1,2,3,\ldots$.
Here, $E_{\Omega}$ is the ground state energy corresponding to $a=0$.

Because the quantum number of each eigenstate is not \textit{a priori} known,
we employ a selection rule based on the symmetry of the system to identify it.
For this purpose, we prepare matrix elements of a proper interpolating operator $\hat{O}_q$ 
with quantum number $q$ as follows 
\be\label{eq:mat_elements}
\langle b|\hat{O}_{q}|a\rangle=\left(U^{{\dagger}}O_{q}U\right)_{ba},
\ee
where $O_q$ is the spin fields representation of $\hat{O}_q$.
Because the model is symmetric under $\mathbb{Z}_2$ transformation of spin fields,  
the possible quantum numbers are $q=\pm 1$, and the following selection rule holds
\be\label{eq:selection_rule}
\langle b|\hat{O}_q|a\rangle \neq 0 \Rightarrow q_bqq_a=1,
\ee
where $q_b,q_a$ are the quantum number of eigenstates $ |b\rangle$ and $|a\rangle$, respectively.
See refs.~\cite{PhysRevD.110.034514,Az-zahra:2024pqa} for a complete derivation of the selection rule.
We choose $| b\rangle =|\Omega\rangle$, which is the ground state with quantum number $q_{\Omega}=+1$.

\section{Tensor network representation and coarse-graining procedure}
\label{sec:coarse_graining}
Transfer matrix has a very large dimension,
making direct diagonalization impractical. 
We therefore employ tensor network technique to obtain the estimation.
Specifically, we represent the transfer matrix in terms of tensor network and
use higher order tensor renormalization group algorithm (HOTRG) \cite{PhysRevB.86.045139}
to reduce its dimension by coarse-graining iteration.
To derive the tensor network representation,
let us recall the partition function in Eq.~(\ref{eq:partition_function_TM}).
First, define a matrix $Y$ 
\begin{equation}\label{eq_y}
Y_{S'k}=\prod_{x=0}^{L_{\rm s}-1}\sum_{k_x=0}^{1}u_{s'_xk_x}\sqrt{\sigma_{k_x}}
\times
\prod_{x=1}^{L_{\rm s}-1}\exp[{\frac{1}{2T}s'_{x+1}s'_x}],
\end{equation}
where $u$ and $\sigma$ are from
EVD of Boltzman weight $\exp[{\frac{1}{T} s'_xs_x}]=\sum_{k_x=0}^{1}u_{s'_xk_x}\sigma_{k_x}u_{k_xs_x}^{\dagger}$, and $k=\{k_x|x=0,1,\ldots, L_{\rm s}-1\}$ is a new defined index.
Thus, the partition function can be written as  
\be
Z=\Tr{(YY^{\dagger})^{L_{\rm t}}}=\Tr{(Y^{\dagger}Y)^{L_{\rm t}}}=\Tr{\mathcal{A}^{L_{\rm t}}}.
\ee
In the last equation, we define a one time slice tensor network $\mathcal{A}\coloneqq Y^{\dagger}Y$,
see Fig.~(\ref{sfig:tensor_network}) for the diagram.
Because $\mathcal{A}$ is composed of the same matrix as $\mathcal{T}$, but with different ordering,
it follows that the EVD of $\mathcal{A}$ is given by 
\be\label{eq:EVD_A}
\mathcal{A}_{kj}=\sum_{a=0}^{2^{L_{\rm s}}-1}W_{ka}\lambda_aW^{\dagger}_{aj}
\ee
where $W$ is the unitary matrix and $\lambda_a$ is the same eigenvalues as in Eq.~(\ref{eq:evd_TM}).
In Fig.~\ref{sfig:tensor_network}, we can see that $\mathcal{A}$ can be obtained by contracting a four-legs tensor $A_{k_xl_xj_xl_{x-1}}$ \cite{PhysRevD.88.056005}
\be
\mathcal{A}_{kj}=\prod_{x=0}^{L_{\rm s}-1}\sum_{l_x=0}^{1}A_{k_xl_xj_xl_{x-1}}.
\ee
This four-legs tensor $A_{k_xl_xj_xl_{x-1}}$ is obtained from \cite{PhysRevD.88.056005}
\be
A_{k_xl_xj_xl_{x-1}}\coloneqq\sqrt{\sigma_{k_x}\sigma_{l_x}\sigma_{j_x}\sigma_{l_{x-1}}}\sum_{s_x=0}^1 \left(u^{\dagger}\right)_{k_xs_x}\left(u^{\dagger}\right)_{l_xs_x}u_{s_xj_x}u_{s_xl_{x-1}}.
\ee
\begin{figure}[t!]
\centering
\begin{subfigure}[b]{0.4\textwidth}
\centering
\includegraphics[width=7cm,height=4cm]{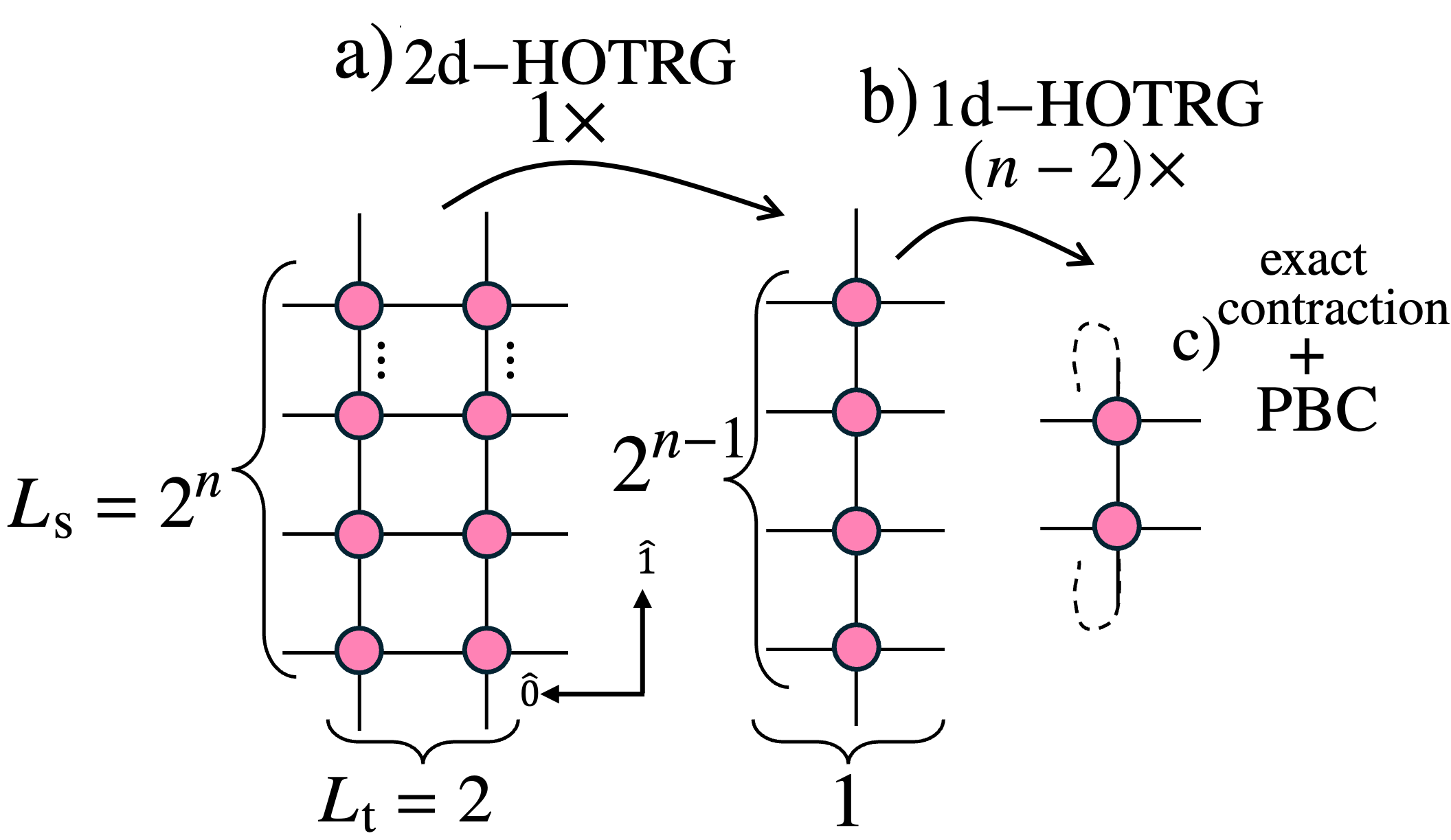}
\caption{}
\label{sfig:tn_pure}
\end{subfigure}\hspace{20mm}
\begin{subfigure}[b]{0.4\textwidth}
\centering
\includegraphics[width=6.2cm,height=4cm]{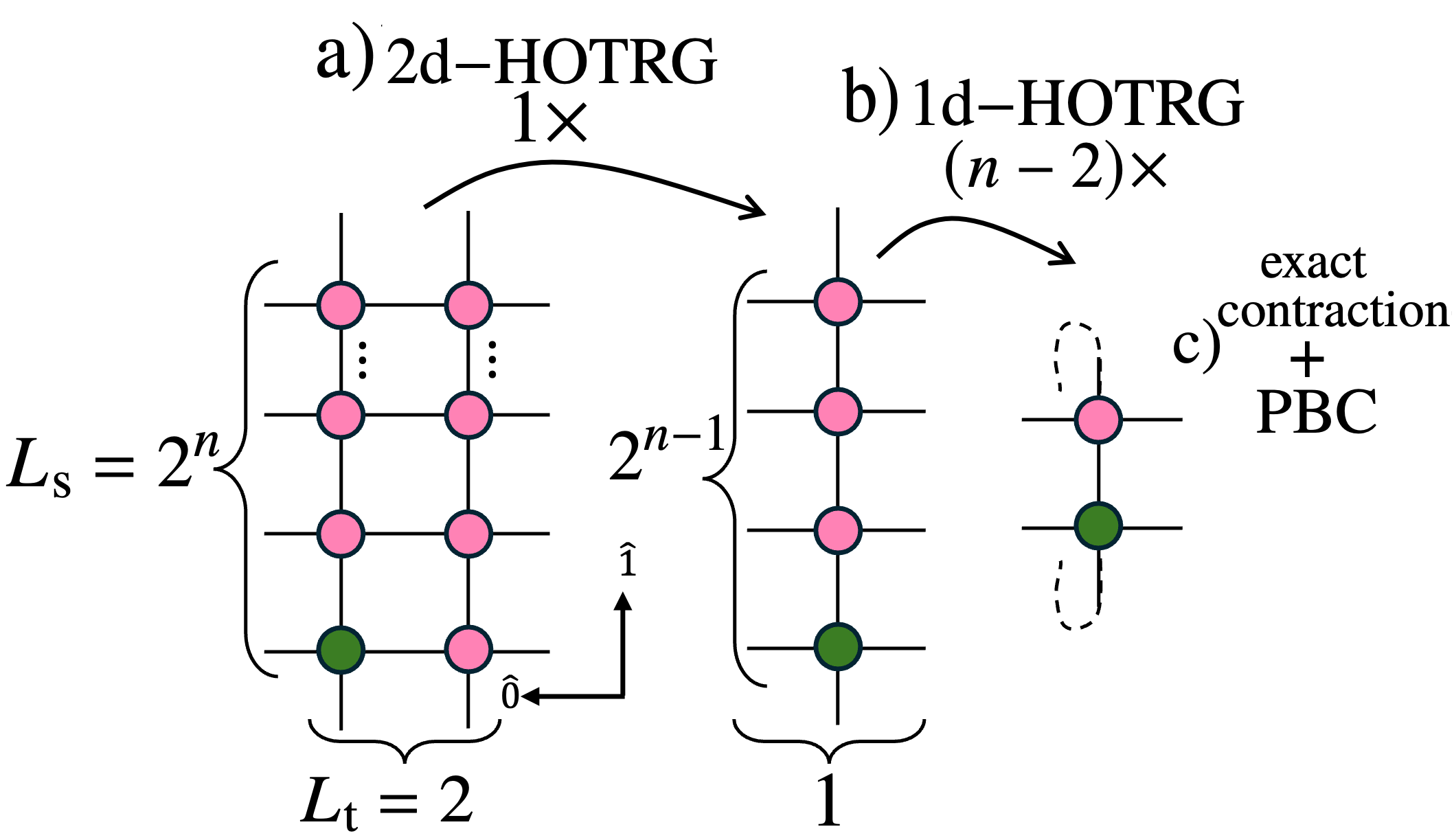}
\caption{}
\label{sfig:tn_impure}
\end{subfigure}
\caption{The coarse-graining procedure for the (a) pure tensor network,  (b) impurity tensor network.
Note that $\hat{0}(\hat{1})$ is the unit vector showing time(space) direction and we show $L_{\rm t}=2$ for readability.}
\label{fig:coarse-graining}
\end{figure}

In \cite{PhysRevD.110.034514}, it is shown that coarse graining of a single--time-slice tensor produces numerical energy spectrum with large errors, 
because the  eigenvalues of $\cal A$ are closely degenerated.
While the square tensor network, $L_{\rm s}=L_{\rm t}$, as used in \cite{PhysRevD.110.034514} yields smaller error for low-lying energy eigenstates, the errors increase drastically for higher excited states. Therefore, in the present work, we modify the computational strategy by computing the spectrum using a tensor network with size $1 < L_{\rm t} < L_{\rm s}$.

The coarse graining for a tensor network with $L_{\rm t} = 2^p$ and $L_{\rm s} = 2^n$, where $p < n$, proceeds as follows:
(a) Apply $p$ coarse-graining iterations to the network in both direction using two-dimensional HOTRG.
(b) Then apply $(n - p - 1)$ coarse-graining iterations in the spatial direction only using one-dimensional HOTRG, resulting in $A^{[n-1]}$.
(c) Finally, by performing an exact contraction of the last two tensors $A^{[n-1]}$ with PBC, we obtain
 \be
 {\cal A}^{L_{\rm t}}_{k_0k_1,j_0j_1}\approx \sum_{l_0,l_1}A^{[n-1]}_{k_0l_0j_0l_1}A^{[n-1]}_{k_1l_1j_1l_0}\eqqcolon {\cal A}^{[n]}_{k_0k_1,j_0j_1},
 \ee
see Fig.~\ref{sfig:tn_pure} for the case $L_{\rm t}=2$.
Subsequently, we apply the EVD to ${\cal A}^{[n]}$, obtaining
 \be
 {\cal A}^{[n]}=\sum_{a}W_a^{[n]}\lambda_a^{[n]}W^{[n]\dagger}
 \ee
 where $W^{[n]}\approx W$ 
 and $\lambda_a^{[n]}\approx \lambda_a^{L_{\rm t}}$ are estimates of unitary matrix and eigenvalues appearing in Eq.~(\ref{eq:EVD_A}), the energy gap then can be estimated as
 \be\label{eq:energy_approx}
 \omega_a\approx \frac{1}{L_{\rm t}}\log\left(\frac{\lambda_{0}^{[n]}}{\lambda_a^{[n]}}\right)\eqqcolon \omega_a^{[\rm hotrg]}\hspace{10mm}
\mbox{ for } a=1,2,3,\ldots.
 \ee
 
 Now we turn to the estimation of matrix elements $\langle \Omega|\hat{O}_q|a\rangle$.
 First, we represent the matrix elements in terms of tensor network quantities, following Ref.~\cite{PhysRevD.110.034514}
 \be\label{eq:mat_element}
 \langle \Omega|\hat{O}_q|a\rangle =\left(\lambda^{-m+1/2}W^{\dagger}\mathcal{A}^{m-1}\mathcal{A}'\mathcal{A}^mW\lambda^{-m-1/2}\right)_{0a}
 \ee
 where $\lambda$ and $W$ are obtained from Eq.~(\ref{eq:EVD_A}), 
 $\mathcal{A}'$ is an impurity tensor network defined as
 \be
\mathcal{A}'_{kj}=\sum_{\{l_x\}}A'_{k_0l_0j_0l_{L_{\rm s}-1}}\prod_{x=1}^{L_{\rm s}-1}A_{k_xl_xj_xl_{l_x-1}}.
 \ee
Here $m$ shows the position of $\mathcal{A}'$ in the network,
where for $L_{\rm t}=2$, $m$ is set to one, see Fig.~\ref{sfig:tn_impure}.
For a single spin field operator put on $x=0$, that is $\hat{O}=s_0$, $A'$ is the impurity tensor defined as
 \be
 A'_{k_0l_0j_0l_{L_{\rm s}-1}}\coloneqq\sqrt{\sigma_{k_0}\sigma_{l_0}\sigma_{j_0}\sigma_{l_{L_{\rm s}-1}}}\sum_{s_0=0}^1 s_0 \left(u^{\dagger}\right)_{k_0s_0}\left(u^{\dagger}\right)_{l_0s_0}u_{s_0j_0}u_{s_0l_{L_{\rm s}-1}}.
 \ee
The coarse-graining of impurity tensor network is performed in the same manner as the pure tensor network case, resulting in $\mathcal{A}^{'[n]}$
 \be
\mathcal{A}^{'[n]}_{k_0k_1,j_0j_1}= \sum_{l_0,l_1}A^{'[n-1]}_{k_0l_0j_0l_1}A^{[n-1]}_{k_1l_1j_1l_0}.
 \ee
By inserting the coarse grained tensor network quantities into Eq.~(\ref{eq:mat_element}),
we obtain an estimate of the matrix elements
\be\label{eq:mat_elemets_approx}
\langle\Omega|\hat{s}_0|a\rangle\approx \left(\left(\lambda^{[n]}\right)^{-(m-1/2)/L_{\rm t}}W^{[n]\dagger}{\cal A}'^{[n]}
W^{[n]}\left(\lambda^{[n]}\right)^{-(m+1/2)/L_{\rm t}}\right)_{0a}
\eqqcolon\langle \Omega|\hat{s}_0|a\rangle^{[\rm hotrg]}.
 \ee

\section{Numerical results}
In this section, we present the results for (1+1)d Ising model in the symmetric phase at $T=2.44$. 
\subsection{Energy spectrum and quantum number classification}
\begin{figure}[t!]
\centering
\begin{subfigure}[b]{0.4\textwidth}
\centering
\includegraphics[width=6.5cm,height=5.5cm]{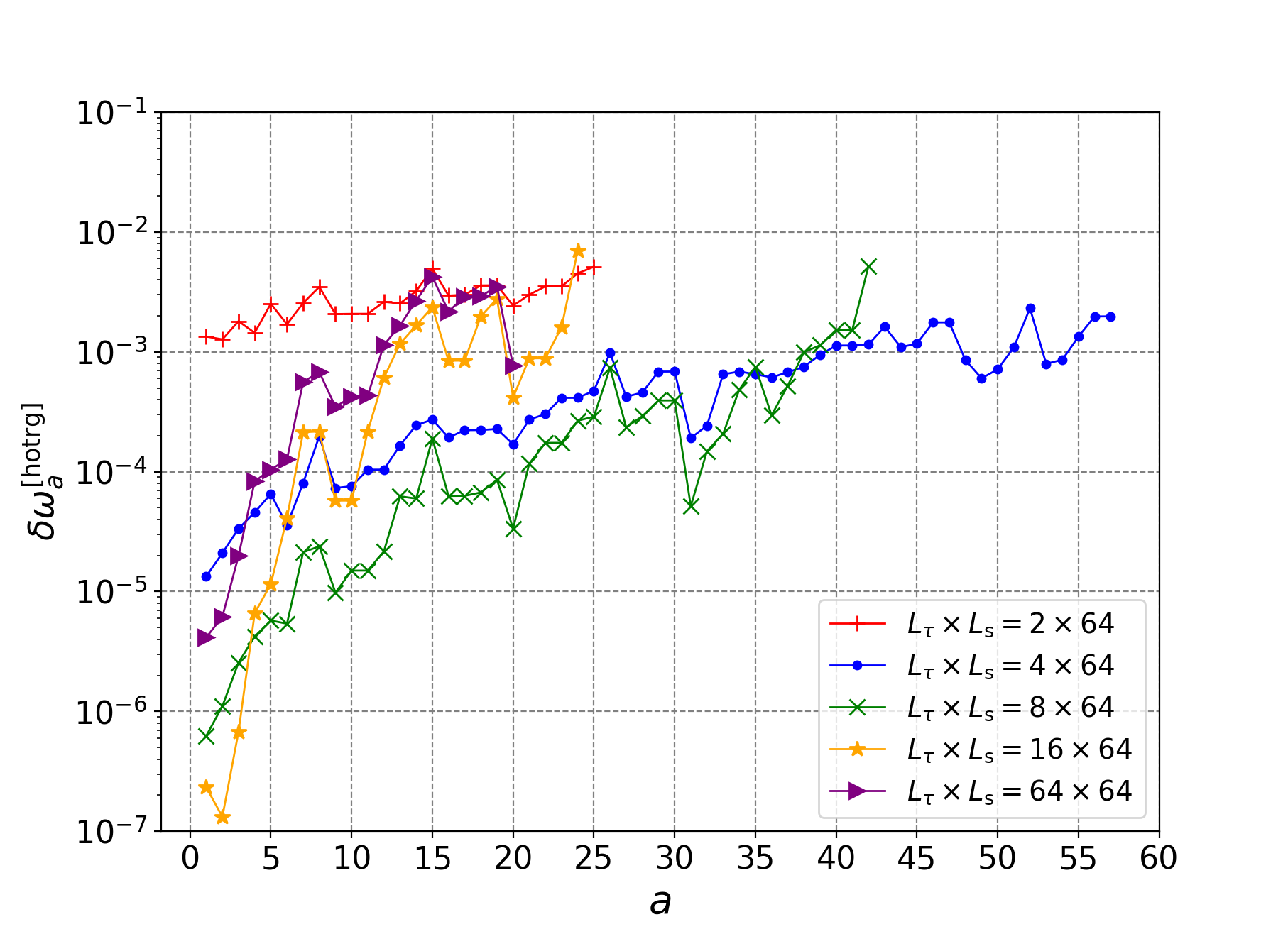}
\caption{}
\label{fig:error}
\end{subfigure}\hspace{10mm}
\begin{subfigure}[b]{0.4\textwidth}
\centering
\includegraphics[width=5.7cm,height=5.5cm]{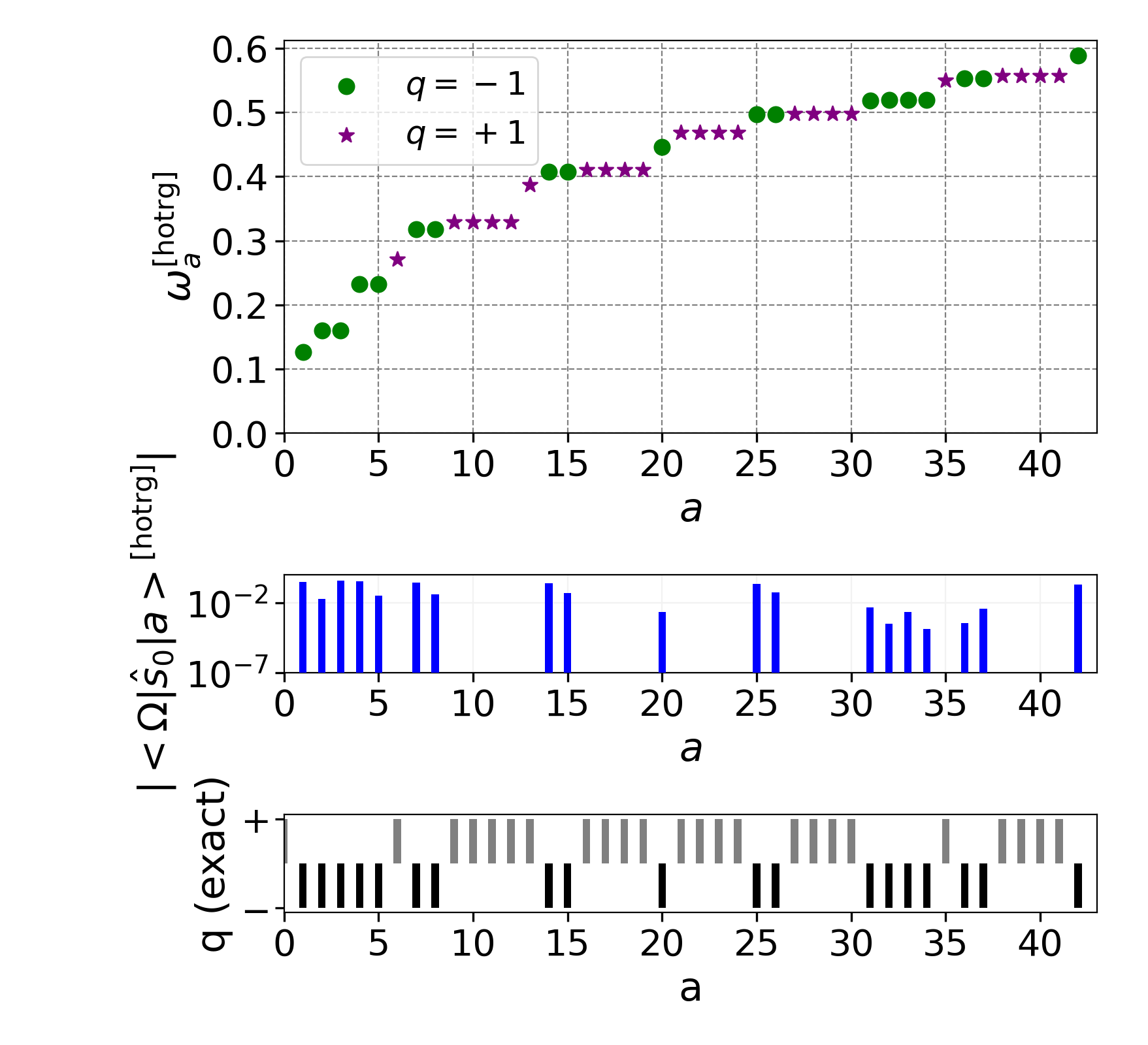}
\caption{}
\label{fig:energy_matelements}
\end{subfigure}
\caption{(a) The relative error of energy spectrum for $L_{\rm s}=64$ computed with $L_{\rm t}=2,4,8,16,64$ using $\chi=80$. 
(b) The numerical energy spectrum, matrix elements for quantum number classification, and exact quantum number for $L_{\rm s}=64$ computed with $\chi=80$ and $L_{\rm t}=8$.}
\label{fig:energy_matelements_error}
\end{figure}
First, we show relative error of the energy 
spectrum $\delta\omega_a=\frac{|\omega_a^{[\rm hotrg]}-\omega_a^{[\rm exact]}|}{|\omega_a^{[\rm exact]}|}$ for $L_{\rm s}=64$ and $L_{\rm t}=2,4,8,16,64$.
The exact results are obtained following Ref.~\cite{PhysRev.76.1232}.
The coarse-graining procedure for $1<L_{\rm t}< L_{\rm s}$ is described in Sec.~\ref{sec:coarse_graining}, while the case $L_{\rm t}=L_{\rm s}$ is discussed in \cite{PhysRevD.110.034514}.
From Fig.~\ref{fig:error}, we see that $L_{\rm t}=8$ gives relatively smaller error compared to the other cases.
The reasons are: the singular values of ${\cal A}^{L_{\rm t}}$ for $L_{\rm t}=8$ are less closely degenerated than those for $L_{\rm t}=2,4$ 
and number coarse graining iteration in time direction is smaller than for $L_{\rm t}=16,64$ which results in fewer source of numerical error.

We compute $\omega_a^{[\rm hotrg]}$ in Eq.~(\ref{eq:energy_approx})
by coarse graining the pure tensor network  with size $L_{\rm t}\times L_{\rm s}=8 \times 64$ using HOTRG
with a cut-off bond dimension $\chi=80$. 
We then identify the corresponding quantum number 
by evaluating $\langle\Omega|\hat{s}_0|a\rangle$.
We coarse grain the impurity tensor network for $\hat{s}_0$ which has the same size as the pure network,
and insert the resulting quantities into Eq.~(\ref{eq:mat_elemets_approx}) to obtain $\langle\Omega|\hat{s}_0|a\rangle^{[\rm hotrg]}$.
Based on the selection rule given in Eq.~(\ref{eq:selection_rule}),
if $|\langle\Omega|\hat{s}_0|a\rangle^{[\rm hotrg]}|\neq 0$, the quantum number of eigenstate $|a\rangle$ is $q=-1$;
otherwise, the quantum number is $q=+1$.
The numerical result of energy with the corresponding quantum number is shown in upper panel of Fig.~\ref{fig:energy_matelements}.
The middle panel shows the results of matrix elements for $a=1-42$. 
We observe non-zero values for the eigenstates number $a=1,2,3,4,5,7,8,14,15,20,25,26,31,32,33,34,36,37,42$, indicating that these states belong to the $q=-1$ sector,
while the remaining states correspond to the $q=+1$ sector.
The exact quantum number, computed following Ref.~\cite{PhysRev.76.1232}, are shown in the bottom panel
which we find the good agreement with the numerical results.
Using $L_{\rm t}=8$,  we are able to compute the energy and identify quantum number up to $a= 42$, which is significantly higher than in Ref.~\cite{PhysRevD.110.034514,Az-zahra:2024pqa}.

\subsection{Momentum and number of particles}
\begin{figure}[t!]
\centering
\begin{subfigure}[b]{0.4\textwidth}
\centering
\includegraphics[width=6cm,height=5cm]{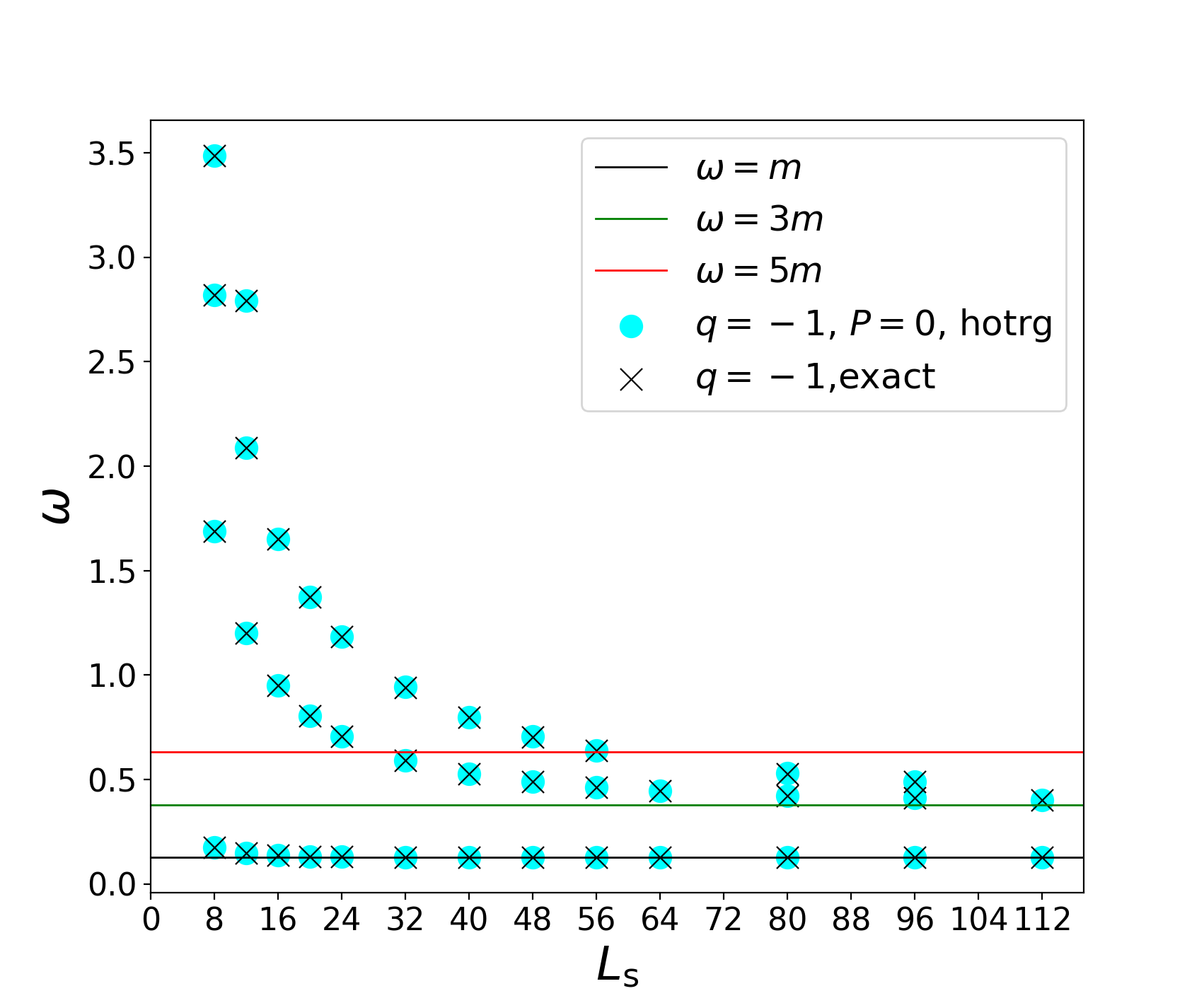}
\caption{}
\label{sfig:qmin}
\end{subfigure}
\begin{subfigure}[b]{0.4\textwidth}
\centering
\includegraphics[width=6cm,height=5cm]{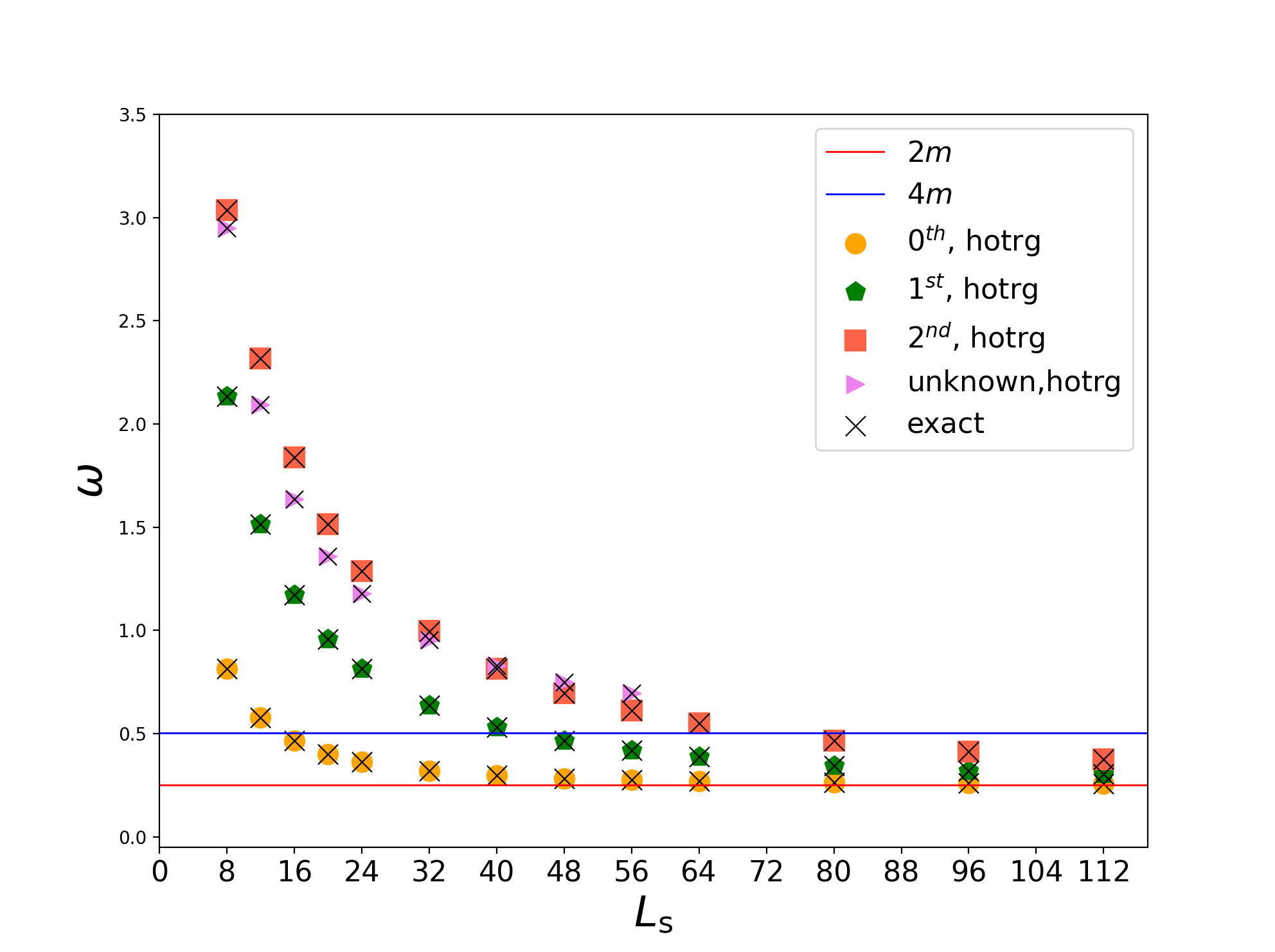}
\caption{}
\label{sfig:qpos}
\end{subfigure}
\caption{The energy spectrum as function of system size computed with $\chi=80$ and $L_{\rm t}=8$
for (a) $q=-1,P=0$ channel,  (b) $q=+1, P=0$ channel.}
\label{fig:spec_over_l}
\end{figure}
To identify the momentum of eigenstates in $q=-1$ sector, we compute the matrix elements $\langle \Omega|\hat{O}_1(P)|a\rangle$
where operator $\hat{O}_1(P)$ is defined as
\be
\hat{O}_1(P)=\frac{1}{L_{\rm s}}\sum_{x=0}^{L_{s}-1}\hat{s}_xe^{-iPx}.
\ee
Here, $P$ is the discrete momentum given by $P=2\pi n/L_{\rm s}$ with $n=0,1,2,\ldots,L_{\rm s}-1$.
The selection rule states that
$\text{if~}\langle \Omega|\hat{O}_1(P)|a\rangle \neq 0$ then 
eigenstates $|a\rangle \text{~has momentum~} P$.
As a first step, we focus on the simplest case that is the identification of eigenstates with momentum $P=0$.
The approximation of the matrix elements, denoted by $\langle \Omega|\hat{O}_1(P)|a\rangle^{[\rm hotrg]}$ is computed in the same manner as matrix elements used to identify quantum number, except that we now coarse-grain the impurity tensor network corresponding to the operator $\hat{O}_1(P)$. 
By examining the non-zero values of $\langle \Omega|\hat{O}_1(P)|a\rangle^{[\rm hotrg]}$ for $L_{\rm s}=64$,
we find that the momentum of eigenstates $a=1,20$ in $q=-1$ sector is $P=0$.
We apply the same procedure for other system sizes $L_{\rm s}=8-112$. 
We then plot the energies from $q=-1,P=0$ sector as function of $L_{\rm s}$, shown in Fig.~\ref{sfig:qmin}.
By examining the system-size dependence of energies,
we observe that the lowest energy level correspond to a one-particle state, since it converges to the rest mass $m$.
The exact value of $m$ at the sufficiently large volume for $T=2.44$ is $m=0.12621870$.
In contrast, the second and third levels are the three-particle states as they approach $3m$ when system size is larger.

Next, we identify the momentum in the $q=+1$ sector using matrix elements of the operator
\be
\hat{O}_2(P,p)=\frac{1}{L_{\rm s}^2}\sum_{x,y=0}^{L_{\rm s}-1}\hat{s}_x\hat{s}_ye^{-ip_1x}e^{-ip_2y},
\ee
where $p_{j}$ for $j=1,2$ are the discrete momenta given by $p_j=2\pi n_j/L_{\rm s}$ with $n_j=0,1,\ldots,L_{\rm s}-1$.  Here $P=p_1+p_2$ is the total momentum and $p=(p_1-p_2)/2$ is the relative momentum.
The selection rules for this case states that, 
for a fixed $P$ irrespective of $p$, if $\langle\Omega|\hat{O}_2(P,p)|a\rangle$ has non-zero values, then the eigenstate $|a\rangle$ has total momentum $P$.
To compute $\langle \Omega|\hat{O}_2(P,p)|a\rangle^{[\rm hotrg]}$,
we coarse grain the impurity tensor network corresponding to the operator $\hat{O}_2(P,p)$.
As in $q=-1$ case, we focus on identifying states with zero total momentum, $P=0$.
From the non-zero values of $\langle \Omega |\hat{O}_2(P,p)|a\rangle^{[\rm hotrg]}$
computed with $L_{\rm t}=8$ and $\chi=80$, 
we determine that, for $L_{\rm s}=64$, the eigenstates in 
the $q=+1$ sector with total momentum $P=0$ are $a=6,13,35$.
We perform this identification for other system sizes as well
and plot energies in the $q=+1, P=0$ sector as a function of $L_{\rm s}$ in Fig.~\ref{sfig:qpos}.
From this figure, we judge that the three lowest energy levels (shown by orange, green and red data points) are the ground, first and the second excited two-particle states because the values approach $2m$ in large system size.
However, the number of particles corresponding to the energy level indicated by the violet triangles cannot be clearly identified, because this energy level is obtained only for relatively small system sizes up to $L_{\rm s} = 56$.
This identification is based not only on the size dependence of the energy, but also on the shape of the corresponding wave function in the $q = +1, P = 0$ sector, which will be discussed in the next section.
\begin{figure}[t!]
\centering
\begin{subfigure}[b]{0.4\textwidth}
\centering
\includegraphics[width=6cm,height=4.8cm]{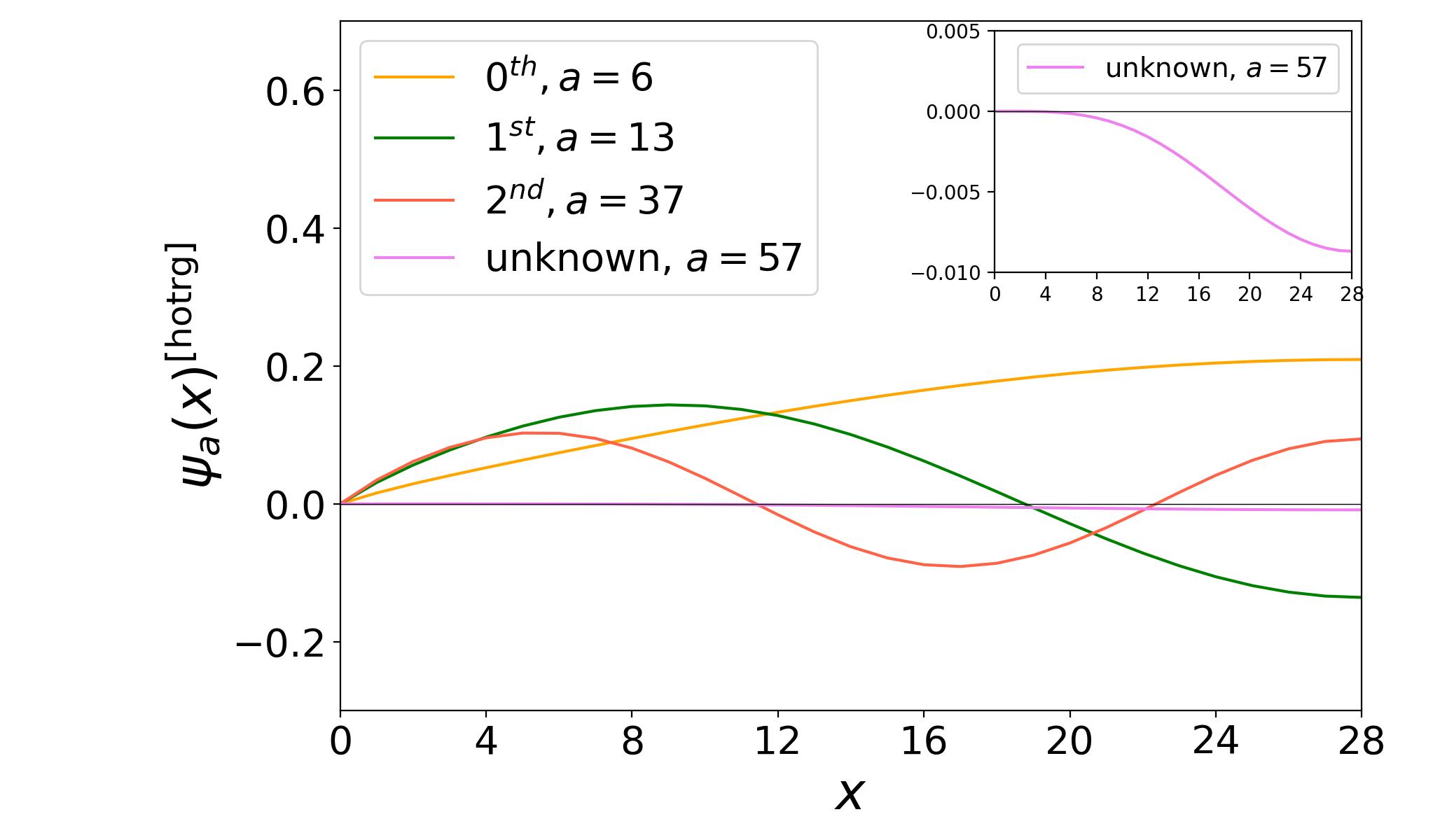}
\caption{}
\label{sfig:wf2f_l56}
\end{subfigure}\hspace{10mm}
\begin{subfigure}[b]{0.4\textwidth}
\centering
\includegraphics[width=6cm,height=4.8cm]{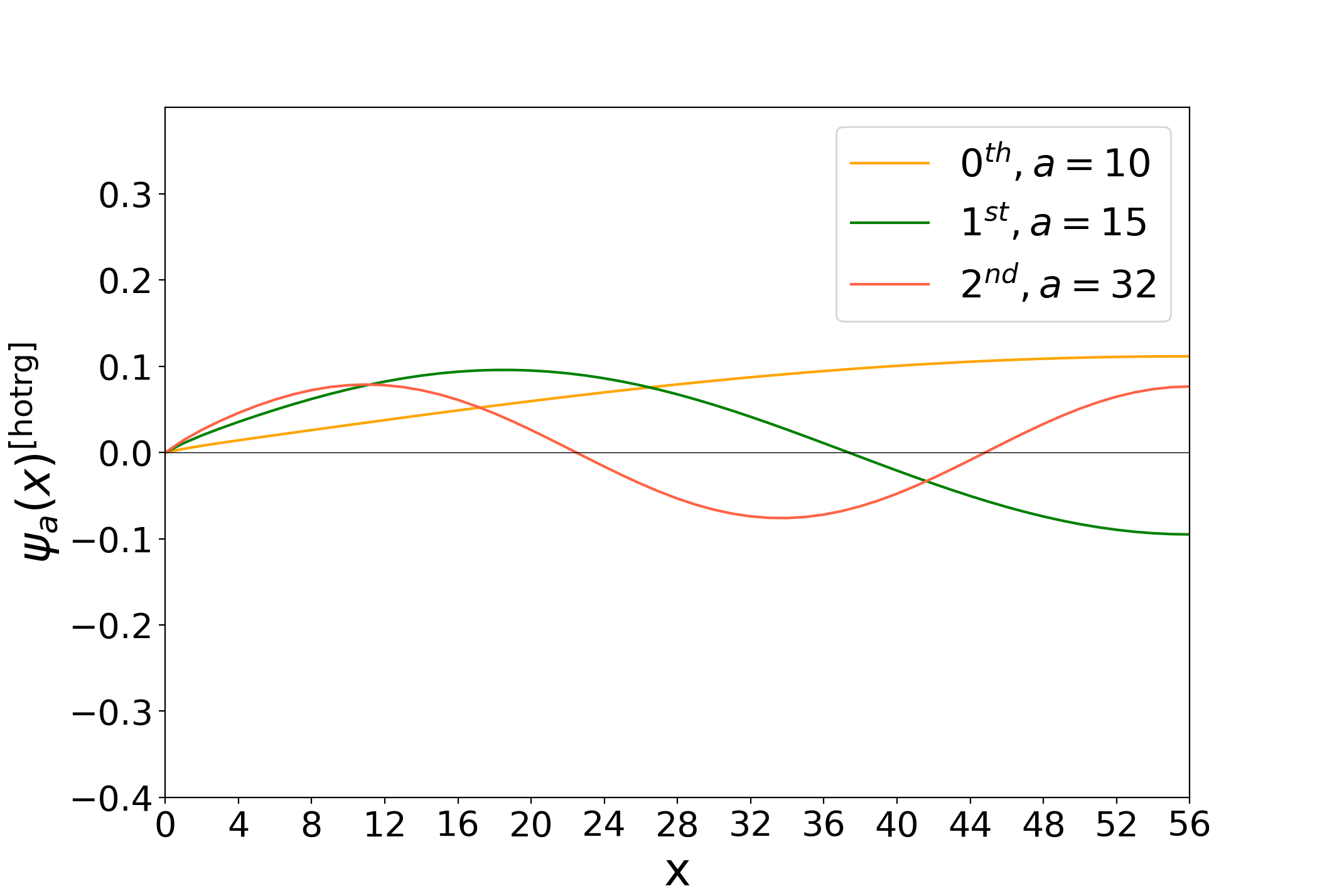}
\caption{}
\label{sfig:wf2f}
\end{subfigure}
\caption{The wave function for $q=+1, P=0$ sector,
computed with $\chi=80, L_{\rm t}=8$ for 
(a) $L_{\rm s}=56$,
(b) $L_{\rm s}=112$.}
\label{fig:wf2f_l112_l56}
\end{figure}
\subsection{Wave function}
The wave function for the $q=+1$, $P=0$ sector is computed from matrix elements
\be
\psi_a(x)=\langle \Omega|\hat{O}_2(x)|a\rangle,
\ee
where $\hat{O}_2(x)=\frac{1}{L_{\rm s}}\sum_{x'}\hat{s}_{x'}\hat{s}_{x'+x}$.
As before,
we coarse-grain the tensor network corresponding to the operator $\hat{O}_2(x)$ and use Eq.~(\ref{eq:mat_elemets_approx})
to obtain the estimate of wave function $\psi_a(x)^{[\rm hotrg]}\coloneqq \langle \Omega|\hat{O}_2(x)|a\rangle^{[\rm hotrg]}\approx \langle \Omega|\hat{O}_2(x)|a\rangle$.
In Fig.~\ref{fig:wf2f_l112_l56}, 
we present the wave function for $L_{\rm s}=56$, 
which is the largest size for which the unknown energy level
shown in Fig.~\ref{sfig:qpos} can be obtained. 
We also include result for $L_{\rm s}=112$, which is the largest size accessible with bond-dimension $\chi=80$.
For $L_{\rm s}=56$, the wave function of the two-particle states from the ground state to the second excited state,
correspond to eigenstates number $a=6,13,37$, respectively. 
The colors in Fig.~\ref{fig:wf2f_l112_l56} are chosen to match those of the corresponding energy levels shown in Fig.~\ref{sfig:qpos}.
The inset of Fig.~\ref{sfig:wf2f_l56} shows an enlarged view of the wave function of the unknown state. 
We observe that the shape is apparently different from the two-particle state wave function and the amplitude is significantly smaller.
Furthermore, for $L_{\rm s}=112$, we get the ground to the second excited two-particle state wave function with $P=0$ which correspond to the eigenstates number $a=10,15,32$, respectively.

\subsection{Two-particle scattering phase shift}
\begin{figure}[t!]
\centering
\begin{subfigure}[b]{0.4\textwidth}
\centering
\includegraphics[width=5.5cm,height=4cm]{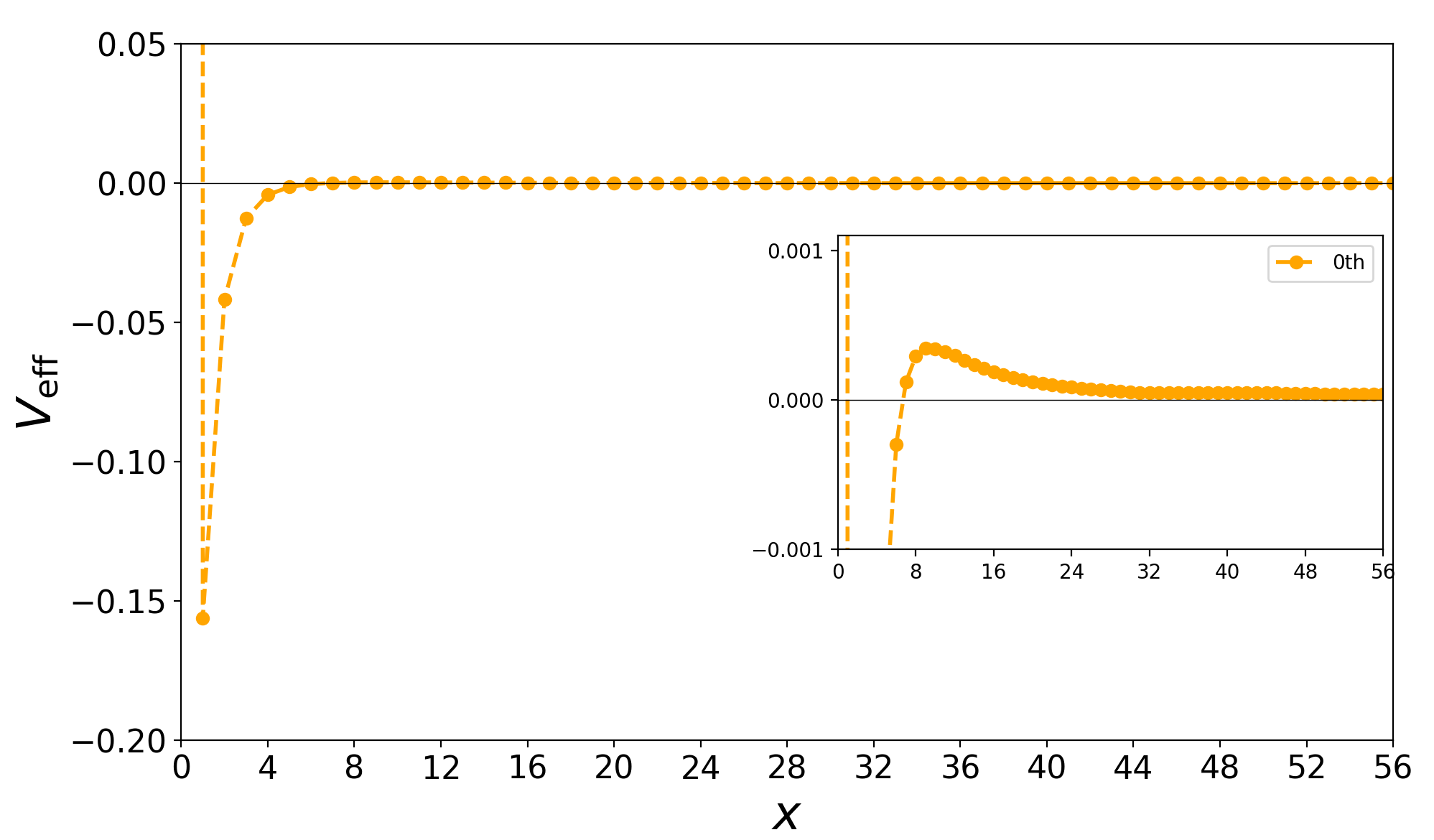}
\caption{}
\label{sfig:pot}
\end{subfigure}
\begin{subfigure}[b]{0.4\textwidth}
\centering
\includegraphics[width=8cm,height=4cm]{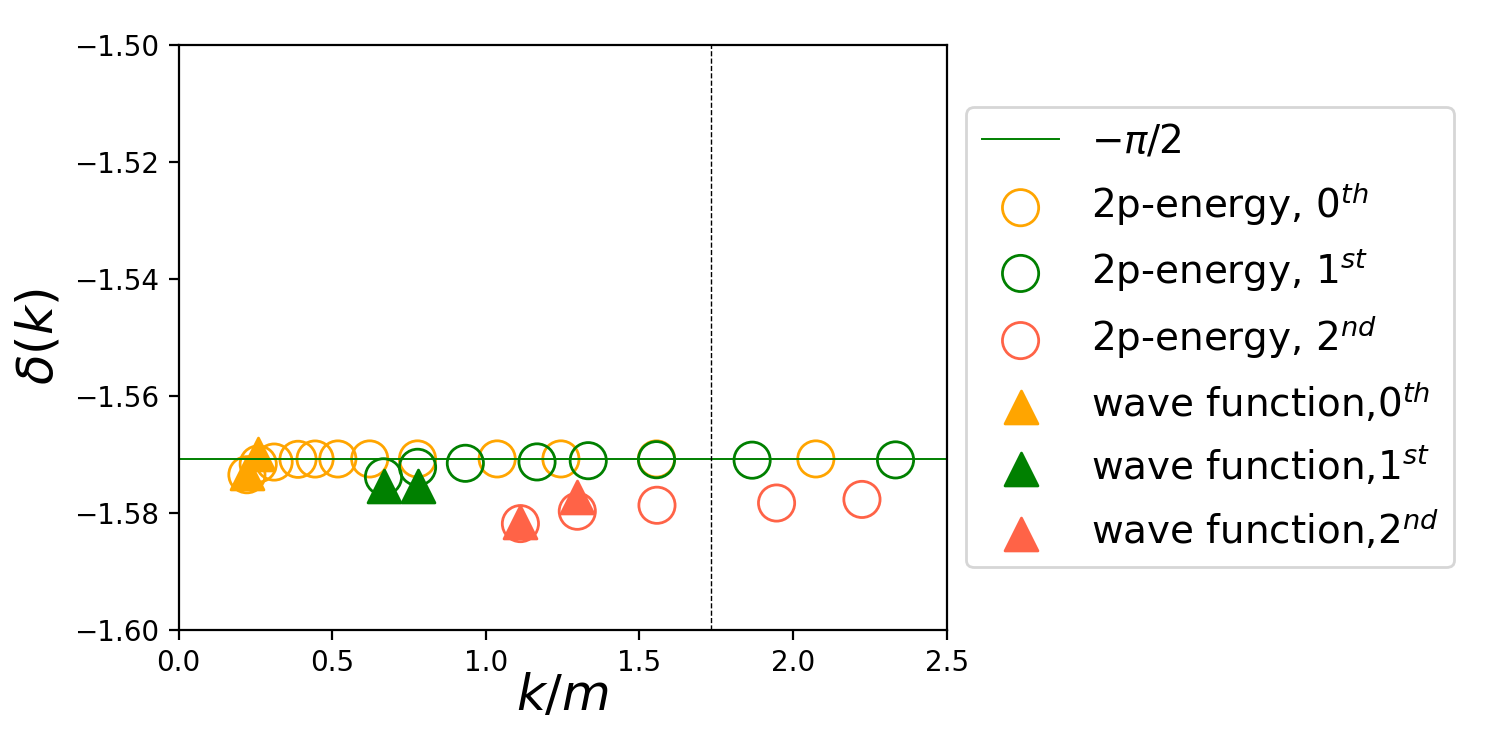}
\caption{}
\label{sfig:ps}
\end{subfigure}
\caption{(a) The effective potential of the ground state wave function for $L_{\rm s}=112$.
(b) The scattering phase shift from L\"uscher's formula (empty circles) and from fitting of the wave function (filled triangles) computed with $\chi=80$ and $L_{\rm t}=8$.}
\label{fig:ps_pot}
\end{figure}
Lastly, we compute the two-particle scattering phase shift using finite volume energy the wave function approach.
First, we show the computation of phase shift from the finite volume energy in center of mass frame ($P=0$) using the L\"uscher's formula \cite{Luscher:1986pf}  
\be
\delta(k) + kL_{\rm s}/2 = 0\mod \pi n\hspace{10mm}\text{for~~}n\in\mathbb{Z}
\ee 
where $k$ is the relative momentum of the two-particle state, obtained from lattice dispersion relation
$
k =\cos ^{-1}(\cosh m +1 -\cosh \omega/2)
$, with $\omega_a^{[\rm hotrg]}$ as the input energy.

Next we describe how to compute the phase shift by fitting the two-particle state wave function $\psi(x)^{[\rm hotrg]}$ with the free wave function,  which is defined as
$
\psi(x)^{[\rm free]}=
A \cos (k(x+ L_{\rm s}/2))$ for $x>R$, where $R$ denotes the effective interaction range, defined as the region beyond which the effective potential vanishes. The effective potential is defined as
\be
V_{\rm eff}(x)= \frac{(\partial^{2})^{\rm lat}\psi(x)}{\psi(x)} +k_{\rm lat}^2
\ee
where $(\partial^{2})^{\rm lat}\psi(x)=\psi(x+1)-2\psi(x)+\psi(x-1)$,
and $\psi(x)^{[\rm hotrg]}$ is taken as the input wave function. Note that $k_{\rm lat}^2=2(1-\cos{k})$, where $k$ is from the lattice dispersion relation.
From Fig.~\ref{sfig:pot}, we observe that for $T=2.44$, $V_{\rm eff}(x)\to 0$ when $R\approx 40$.
Since the fitting method is only applicable when the condition $L_{\rm s}/2 >R$ is satisfied,
we perform the wave function fitting only for $L_{\rm s}=96,112$.
The fitting parameters are $A,k$ and the phase shift is extracted as $\delta(k)=kL_{\rm s}/2 \mod{n\pi}$. 
In Fig.~\ref{sfig:ps}, we observe good agreement between the phase shift obtained from the wave function fitting and the finite-volume energy method, as well as with the exact prediction $\delta(k)^{[\rm exact]}=-\pi/2$
\cite{Gattringer:1992np}.

\section{Summary}
We have investigated of the multi-particle states of the (1+1)d Ising model
using a spectroscopy scheme based on tensor renormalization group method.
Our results showed that tensor network with $L_{\rm t}=8$ yields energy spectrum with relatively smaller errors,
enabling reliable determination of higher excited state energies and their corresponding quantum numbers.
In particular, we correctly identified eigenstates with zero total momentum in $q=\pm1$ sectors.
From the system size dependence of the energy levels, we identified one- and three-particle states in the $q=-1,P=0$ sector,
and two-particle states in the $q=+1,P=0$ sector.
Next, we computed the two-particle state wave function using the impurity tensor network method, 
and successfully extract the two-particle state wave function from the ground to the second excited state.
Finally, we extracted the two-particle scattering phase shift using L\"uscher's formula and a fitting procedure applied to the wave function
outside the interaction range $R\approx 40$ for $T=2.44$.
The phase shift obtained from both methods are consistent
with each other and agree with the exact result \cite{Gattringer:1992np}.
As future work, we plan to analyze the three-particle states in detail and apply this scheme to other quantum field theories.
\acknowledgments
F.I.A.   is supported by JST SPRING, Grant No. JPMJSP2135.
S.T. is supported in part by JSPS KAKENHI Grants No.~21K03531, No.~22H05251, and No.~25K07280.
T.Y. is supported in part by JSPS KAKENHI Grant No.~23H01195 and No.~23K25891 , and MEXT as ``Program for Promoting Researchers on the Supercomputer Fugaku'' (Grant No. JPMXP1020230409).


\begin{thebibliography}{99}
\bibitem{Luscher:1986pf}    
M. Luscher, \emph{Volume Dependence of the Energy Spectrum in Massive Quantum Field Theories. 2. Scattering States},  
\href{https://doi.org/10.1007/BF01211097}
    {\emph{ Commun. Math. Phys.} \textbf{105} (1986) 153--188}.
    
\bibitem{Itou:2023img} 
E. Itou. and A. Matsumoto, and Y. Tanizaki, \emph{Calculating composite-particle spectra in Hamiltonian formalism and demonstration in 2-flavor QED$_1+1d$},\href{https://doi.org/10.1007/JHEP11(2023)231}
    {\emph{ JHEP} \textbf{11} (2023) 231} {[\href{https://arxiv.org/abs/2307.16655}{\emph{\tt  2307.16655}}]}.
    
\bibitem{Itou:2024psm}
 E. Itou. and A. Matsumoto, and Y. Tanizaki, \emph{DMRG study of the theta-dependent mass spectrum in the 2-flavor Schwinger model},\href{https://doi.org/10.1007/JHEP09(2024)155}
    {\emph{ JHEP} \textbf{09} (2024) 155} {[\href{https://doi.org/10.48550/arXiv.2407.11391}{\tt  2407.11391}]}.
 
 \bibitem{PhysRevD.110.034514}
F. I. Az-zahra, S. Takeda, T. Yamazaki,  \emph{Spectroscopy with the tensor renormalization group method}, \href{https://doi.org/10.1103/PhysRevD.110.034514}
    {\emph{\tt Phys. Rev. D} \textbf{110} (2024) 034514} {[\href{https://doi.org/10.48550/arXiv.2404.15666}{\tt 2404.15666}]}.
    
 \bibitem{Az-zahra:2024pqa}
F. I. Az-zahra, S. Takeda, T. Yamazaki,  \emph{Spectroscopy using tensor renormalization group method}, \href{https://doi.org/10.48550/arXiv.2411.19437}
    {\emph{\tt 2411.19437} \textbf{}}{}.    
    
\bibitem{PhysRevB.86.045139}
Z. Y. Xie, J. Chen, M. P. Qin,  J. W. Zhu, L. P. Yang, T. Xiang, \emph{Coarse-graining renormalization by higher-order singular value decomposition},\href{https://link.aps.org/doi/10.1103/PhysRevB.86.045139}
    {\emph{ Phys. Rev. B} \textbf{86} (2012) 045139} {[\href{https://arxiv.org/abs/1201.1144} {\emph{\tt 1201.1144}}]}.
   
 \bibitem{PhysRevD.88.056005}
Y. Liu, Y. Meurice, M. P. Qin, J. Unmuth-Yockey, T. Xiang, Z. Y. Xie., J. F. Yu. H. Zou , \emph{Exact blocking formulas for spin and gauge models}, \href{https://link.aps.org/doi/10.1103/PhysRevD.88.056005}
    {\emph{ Phys. Rev. D} \textbf{88} (2013) 056005 } {[\href{
https://doi.org/10.48550/arXiv.1307.6543} {{\tt 1307.6543}}]}.

\bibitem{PhysRev.76.1232}
B. Kaufman, \emph{Crystal statistics. II. partition function evaluated by spinor analysis},\href{https://link.aps.org/doi/10.1103/PhysRev.76.1232}
    {\emph{ Phys. Rev. } \textbf{76} (1946) 1232}.
    
\bibitem{Gattringer:1992np}
C. R. Gattringer,  and C. B Lang, \emph{Resonance scattering phase shifts in a 2-d lattice model}, \href{https://doi.org/10.1016/0550-3213(93)90155-I}
    {\emph{ Nucl. Phys. B} \textbf{391} (1993)} {[\href{https://arxiv.org/abs/hep-lat/9206004} {{\tt hep-lat/9206004}}]}.    
\end{thebibliography}
\end{document}